# Negative strange-quark chemical potential
# A concise, well-defined indication of the quark-deconfinement phase transition

Apostolos D. Panagiotou, Panayotis Katsas, Elpida Gerodimou

*University of Athens, Department of Physics*
*Nuclear & Particle Physics Division*
*Panepistimiopolis, GR-157 71 Ilissia, Athens, Hellas*

## Abstract

We have studied the variation of a thermodynamic quantity, the strange-quark chemical potential, in the phase diagramme of nuclear matter, by employing the partition function in each domain and enforcing strangeness conservation. We propose that the change in the sign of the strange-quark chemical potential, from positive in the hadronic phase to negative in the deconfined quark-gluon phase, to be a new, unique, concise and well-defined indication of the quark-deconfinement phase transition in nuclear matter. This signature is independent of model parameters and interaction mechanisms. We propose also that, for a state in the deconfined region following an isentropic path to hadronization via a second order phase transition, the fugacities of the equilibrated quark flavours, once fixed in the primordial state, remain constant throughout the hadronization process. This would enable the knowledge of the thermodynamic quantities of states situated beyond the hadronic phase. Data from nucleus-nucleus interactions at AGS and SPS give support to our proposals.

It is generally expected that ultra-relativistic nucleus-nucleus collisions will provide the basis for strong interaction thermodynamics of nuclear and hadronic matter. In normal matter, colour confinement prohibits the quarks from ever escaping the hadronic 'bag' and be free outside. However, under certain conditions of high temperature, prevailing in the very early moments of the cosmos, or density, existing in the core of neutron stars, both now approached in ultra-relativistic nucleus-nucleus interactions in the laboratory, we expect strongly interacting hadronic matter to undergo two phase transitions beyond the Hadronic Gas (HG) phase: quark deconfinement, giving rise to a continuous QCD state of (initially) constituent-mass quarks and gluons in an extended volume of nuclear size, and chiral symmetry restoration, denoting the diminishing of the quark mass to current-mass.

At finite baryon density, lattice calculations suggest that deconfinement should take place earlier, followed by the weakening of the q-q correlation-interaction[1] and the decrease of the quark mass with increasing temperature, thus defining the Deconfined Quark Matter (DQM) domain. Full chiral symmetry restoration is achieved asymptotically with the formation of the ideal Quark – Gluon Plasma (QGP) of non-interacting, current-mass quarks. We may thus define a phase diagramme of strongly interacting matter at finite baryon density with three regions: HG – DQM – (ideal) QGP [1-3]

Such a 3-region phase diagramme should be described by a systematic variation of thermodynamic quantities, as one proceeds from one domain to the other. The task is to establish a discrete and well-defined quantity, which changes concisely and observably as nuclear matter changes phase. We propose and show that the strange-quark chemical potential, $\mu_s$, created in an interacting system with finite baryon number density, is the sought-for thermodynamic quantity. It is a unique, concise and well-defined observable-signature, whose behaviour in the various phases can be precisely known, derived from the corresponding Equation of State (EoS).

Many observables-signatures of the formation of deconfined matter have been proposed and searched for, including the suppression of J/ψ [4], the broadening and shift of hadronic resonance mass [5], the enhancement of strangeness [6], the formation of disoriented chiral condensates (DCC) [7]. These theoretical observables are being tested and compared against experimental data from nucleus-nucleus interactions at the SPS, however without conclusive results so far [8-10], due to known (and unknown) interfering effects, stemming from the various possible interaction mechanisms, statistical fluctuations of basic quantities measured, as well as model dependence and model uncertainties. On the other hand, the change in the sign of the strange-quark chemical potential, a distinct and well-defined characteristic thermodynamic quantity in each phase, may provide an unambiguous, unique signature of the deconfinement phase transition, which is independent of models and interaction mechanisms. In the following we shall derive and discuss the EoS and the functional form of the strange-quark chemical potential in each of the three regions of the phase diagramme.

---

[1] Lattice calculations show that $\alpha_s$ (T=300 MeV) = 0.3 and that $\varepsilon/T^4$ (T ~ $2T_d$ ~ 350 MeV) ~ $0.85\varepsilon_{SB}$, $3P/T^4$ (T ~ $2T_d$) ~ $0.65\varepsilon_{SB}$, where $\varepsilon_{SB}$ is the Stefan-Boltzmann ideal gas limit [31].

In the HG phase, the hadron-mass spectrum is given by the partition function, $\ln Z_{HG}$, in the Boltzmann approximation. We assume that the hadronic state has attained thermal and chemical equilibration of three quark flavors (u, d, s):

$$\ln Z_{HG}(T,V,\lambda_q,\lambda_s) = Z_m + (\lambda_q^3 + \lambda_q^{-3})Z_n$$
$$+ Z_K(\lambda_q \lambda_s^{-1} + \lambda_s \lambda_q^{-1})$$
$$+ Z_Y(\lambda_s \lambda_q^2 + \lambda_q^{-2}\lambda_s^{-1})$$
$$+ Z_\Xi(\lambda_s^2 \lambda_q + \lambda_s^{-2}\lambda_q^{-1})$$
$$+ Z_\Omega(\lambda_s^3 + \lambda_s^{-3}) \qquad (1)$$

The one particle Boltzmann partition function $Z_k$ is

$$Z_k(V,T) = \frac{VT^3}{2\pi^2} \sum_j g_j \left(\frac{m_j}{T}\right)^2 K_2\left(\frac{m_j}{T}\right)$$

where the summation runs over the resonances of each hadron species with mass $m_j$ and $g_j$, the degeneracy factor, counts the spin and isospin degrees of freedom of the j-resonance. $\lambda_i = \exp(\mu_i/T)$ is the fugacity of the i-quark, which controls the quark content of the k-hadron, $\mu_i$ being the quark chemical potential (i=u,d,s). Hadron species with mass up to 2.5 GeV/c$^2$ have been used.

Strangeness conservation in strong interactions necessitates:

$$<N_s - N_{\bar{s}}> = \frac{T}{V}\frac{\partial}{\partial \mu_s}\left[\ln Z_{HG}(V,T,\lambda_s,\lambda_q)\right] = 0$$

which reduces to

$$Z_K(\lambda_s\lambda_q^{-1} - \lambda_q\lambda_s^{-1}) + Z_Y(\lambda_s\lambda_q^2 - \lambda_q^{-2}\lambda_s^{-1}) + 2Z_\Xi(\lambda_s^2\lambda_q - \lambda_s^{-2}\lambda_q^{-1}) + 3Z_\Omega(\lambda_s^3 - \lambda_s^{-3}) = 0 \qquad (2)$$

This is an important equation, as it defines the relation between the temperature and the light-, strange-quark fugacities, $\lambda_q$, $\lambda_s$ (q = u, d), hence the relation between the quark chemical potentials $\mu_q$, $\mu_s$ and T in the equilibrated primordial state. For given $\mu_q$, Eq. (2) defines the variation of the strange-quark chemical potential with temperature in the HG phase. Eq. (2) can be used to derive the true (transverse flow velocity-independent) temperature of the equilibrated state, once the fugacities $\lambda_q$ and $\lambda_s$ are known from experimental strange particle yield ratios [1-3].

In the ideal QGP region, the EoS for the non-interacting, current-mass u,d,s-quarks and gluons has the form:

$$\ln Z_{QGP}(T,V,\mu_q,\mu_s) = \frac{V}{T}\left\{\frac{37}{90}\pi^2 T^4 + \mu_q^2 T^2 + \frac{\mu_q^4}{2\pi^2} + \frac{g_s m_s^{0\,2} T^2}{2\pi^2}(\lambda_s + \lambda_s^{-1})K_2\left(\frac{m_s^0}{T}\right)\right\} \qquad (3)$$

where $m_s^o$ is the current-mass of the s-quark.

Strangeness conservation: $(T/V)[\partial \ln Z_{QGP}/\partial\mu_s] = 0$, gives $\lambda_s = \lambda_s^{-1} = 1$, or

$$\mu_s^{QGP}(T,\mu_q) = 0 \qquad (4)$$

independent of temperature.

To describe the deconfined-quark state in the DQM region, we use the following picture: Beyond but near the HG boundary, at T ≥ $T_d$, $T_d$ being the deconfinement temperature, the correlation-interaction between the deconfined quarks is near maximum ($\alpha_s(T) \leq 1$), and the quark-correlations resemble "hadron-like" states. With increasing temperature, this correlation-interaction weakens as colour mobility increases and $\alpha_s \to 0$. The masses of (anti)quarks vary with the temperature of the state and scale according to a prescribed way. The initially constituent-mass of the quarks – at T = $T_d$ – decreases and, as the DQM domain approaches asymptotically the QGP one, as T → $T_{qgp}$, $T_{qgp}$ being the asymptotic temperature in the ideal QGP region, the current-mass of quarks is attained (full chiral symmetry restoration).

The equation of state in the DQM region should lead to the EoS of the hadronic phase, Eq. (1), at $T \leq T_d$ and to the EoS of the ideal QGP, Eq. (3), at $T \sim T_{qgp}$. To construct the EoS in the DQM phase, we use the two order parameters:

(a) The average thermal Wilson loop: $<L> = \exp(-F_q/T) \sim R_d(T) = 0 \to 1$, as $T = T_d \to T_{qgp}$, describing the quark deconfinement and subsequent colour mobility, $F_q$ being the free quark energy.

(b) The scalar quark density: $<\bar{\psi}\psi> \sim R_{ch}(T) = 1 \to 0$, as $T = T_d \to T_{qgp}$, denoting the scaling of the quark mass with temperature.

We assume that above $T_d$ "hadronic language" is still appropriate to some extent and that the quarks have a degree of correlation-interaction resembling "hadron-like" entities, since $1 > \alpha_s > 0$. The diminishing of this correlation-interaction is approximated by the factor $(1-R_d) = 1 \to 0$, as $T = T_d \to T_{qgp}$. Note that effectively $(1-R_d) \sim \alpha_s(T)$ in the DQM region. To account for the "effective mass" of the state as a function of temperature, we assume the mass of the quarks to decrease and reach the current-mass value as $T \to T_{qgp}$. The quark mass scales with temperature as:

$$m_q^* = R_{ch}(T)(m_q - m_q^o) + m_q^o,$$

where $m_q$ and $m_q^o$ are the constituent and current quark masses, respectively (q=u,d,s). Similarly, the 'effective hadron' mass scales as:

$$m_i^* = R_{ch}(T)(m_i - m_i^o) + m_i^o,$$

where $m_i$ is the hadron mass in the hadronic phase and $m_i^o$ is equal to the sum of the current-mass of the hadron's quarks. In the EOS, the former scaling is employed in the mass-scaled partition function in the ideal QGP phase, $\ln Z_{QGP}^*$, whilst the latter in the mass-scaled partition function in the HG phase, $\ln Z_{HG}^*$, which also accounts for the hadron species. Note that this mass-scaling is effectively equivalent to the one given by the Nambu–Jona-Lasinio formalism [11].

Employing the described dynamics, we construct the EoS of the DQM region:

$$\ln Z_{DQM}(V,T,\lambda_q,\lambda_s) = [1-R_d(T)]\ln Z_{HG}^*(V,T,\lambda_q,\lambda_s) + R_d(T)\ln Z_{QGP}^*(V,T,\mu_q,\mu_s) \quad (5)$$

The factor $[1-R_d(T)]$ describes the weakening of the correlation-interaction of the deconfined quarks, forming the "hadron-like" entities and the $\ln Z_{HG}^*$ gives the mass-scaling of these entities with increasing temperature. In the second term, the factor $R_d(T)$ defines the rate of colour mobility, whilst the $\ln Z^*_{QGP}$ represents the state as it proceeds towards the ideal QGP region. Thus, at T = $T_d$, the EoS of the DQM region goes over to the EoS in the HG phase, and at T → $T_{qgp}$, to the EoS in the ideal QGP region.

Strangeness conservation leads to:

$$[1 - R_d(T)] [Z^*_K (\lambda_s \lambda_q^{-1} - \lambda_q \lambda_s^{-1}) + Z^*_Y (\lambda_s \lambda_q^2 - \lambda_s^{-1} \lambda_q^{-2}) + 2Z^*_\Xi (\lambda_s^2 \lambda_q - \lambda_s^{-2} \lambda_q^{-1})$$
$$+ 3Z^*_\Omega (\lambda_s^3 - \lambda_s^{-3})] + R_d(T) g_s m_s^{*2} K_2 \left( \frac{m_s^*}{T} \right) (\lambda_s - \lambda_s^{-1}) = 0 \qquad (6)$$

Eq. (4) defines, for given $\mu_q$, the variation of the strange-quark chemical potential with temperature in the DQM domain (T = $T_d$ → $T_{qgp}$).

Combining Eq's (2,4,6) we obtain the variation of the strange-quark chemical potential with temperature in the entire phase diagramme. Fig. 1 exhibits the well-defined behaviour of $\mu_s(T)$ for given light-quark chemical potential, say $\mu_q$ = 0.45T. $\mu_s$ attains positive values in the HG phase, due to the positive change of the Gibbs free energy, $\mu_s$ = $(\partial G / \partial N_s)|_T$ as the strange-hadron number density, $N_s$, increases with increasing temperature. It approaches zero as the hadron density reaches its asymptotic Hagedorn limit [12] on the end of the hadronic phase at the deconfinement temperature $T_d$, where a phase transition to quark-gluon matter takes place. At this temperature, $\mu_s(T_d)$ = 0. Then, $\mu_s$ grows negative in the DQM phase, due to the negative change of the Gibbs energy of the deconfined-quark – gluon state, $\mu_s$ = $(\partial G / \partial n_s)|_T$, $n_s$ being the strange-quark number density, giving rise to quark mass and interaction energy. Finally it approaches zero as the ideal QGP phase is reached asymptotically. We note that the sign of $\mu_s(T)$ − positive in the HG, negative in the DQM and zero in the ideal QGP domains − is independent of the particular form of the parameters $R_d$, $R_{ch}$ used in the EoS and unique in each region. A detailed, quantitative treatment of the EoS in the DQM phase from 'first principles' will require the use of a three-flavour effective Lagrangian in the Nambu – Jona-Lasinio formalism [work in progress].

Systematic experimental studies at relativistic energies, aiming at the understanding of nucleus-nucleus interactions and the possibility of producing the deconfined partonic phase, have been performed at the AGS accelerator at BNL and the SPS at CERN, spanning the energy range between 11 and 200 GeV per nucleon. The data obtained by the experiments E802 [13,14] and NA35 [15,16], NA49 [17-19] at AGS and SPS, respectively, have been analyzed in terms of several statistical-thermal models: The Statistical Bootstrap

Model [12], as developed with the inclusion of the strange quantum number and isospin asymmetry, the so-called SSBM [20-23], as well as others employing the canonical and grand-canonical formalisms [24-26]. In Table 1 we summarize the results of these analysis for several nucleus-nucleus interactions, from which the thermodynamic quantities T, $\mu_q$ and $\mu_s$ have been deduced.

Fig. 2 is a plot of the mean values of the temperature and strange-quark chemical potential, obtained from these calculations, Table 1. We observe that the interactions Si+Au at 14.6 AGeV, Au+Au at 11.6 AGeV and Pb+Pb at 40 and 158 AGeV have positive $\mu_s$, whilst the interactions S+S and S+Ag, both at 200 AGeV, exhibit negative values[2]. This is the first experimental confirmation of negative values for the strange-quark chemical potential.

In contrast to all other thermal-statistical models, the SSBM [20,21] incorporates the hadronic interactions in the EoS through the bootstrap equation and thus gives, in a definitive way, the limits of the hadronic phase as a result of the branch point of this equation. The borderline of the hadronic phase with zero strangeness is the projection on the (T, $\mu_q$) plane of the intersection of the 3-dimentional (T, $\mu_q$, $\mu_s$) critical surface of the bootstrap equation and the strangeness conservation surface. The analysis with this model has shown that the S+S interaction is situated mostly (75%) outside the hadronic phase, whilst the S+Ag is on the deconfinement line. In addition, the analysis has identified a large (~ 30%) entropy enhancement of the experimental data compared to the model, which effect was also observed by other calculations [24,26]. This enhancement may be attributed to contributions from the DQM phase with the many more partonic degrees of freedom. For the Pb+Pb interaction, the SSBM analysis [27] has shown that this system is located well within the hadronic phase. In addition, the SSBM and the other thermal models do not find any entropy enhancement for this interaction. All these results are corroborating the observations of positive strange-quark chemical potential for the Pb+Pb and negative for the sulfur-induced interactions, positioning the former within and the later beyond the HG domain, in the DQM phase. Fig. 3 shows the phase diagramme with the SSBM maximally extended[3] deconfinement line, as well as the location (average T, $\mu_q$ values) of the equilibrated states of several nucleus-nucleus interactions, deduced from the analysis of experimental data with thermal models, Table 1. We observe that only the sulfur-induced interactions at 200 AGeV, having negative $\mu_s$, are situated beyond the deconfinement line, whilst all other interactions with positive $\mu_s$ are located well within the hadronic phase (see fig. 2).

---

[2] The T, $\mu_q$, $\mu_s$ values for the sulfur-induced interactions were obtained from fits to particle yields with thermal hadronic models [24-26], which cannot distinguish the onset of the DQM phase. However, beyond but near the deconfinement line, the state may still be assumed 'hadron-like' and hadronic models may describe the state adequately.

[3] The maximally extended HG phase (deconfinement line) is defined for $T_0$ = 183 MeV (corresponding to $B^{1/4}$ = 235 MeV) which is the maximum temperature at $\mu_q$ = 0 for non-negative strange-quark chemical potential in the HG domain. This $T_0$ will decrease slightly if charm-hadron resonances are included in the mass spectrum. Recent lattice QCD calculations give $T_0$ ~ 175 MeV for three quark flavours [31].

Fig. 4 shows the theoretical variation of the strange-quark chemical potential throughout the phase diagramme, obtained from Eq's (2,4,6) for $\mu_q$ = 82 MeV, together with the mean T, $\mu_s$ values given by thermal model fits to the sulfur-induced interactions. In this figure there is no intention to show a fit to the points, since the EoS used in the DQM region is constructed with the empirical dynamics of the deconfined region without adjusting any parameters (see also footnote **2**), but rather to exhibit the qualitative correspondence between the experimentally deduced thermodynamic quantities T, $\mu_s$ of the (considered as deconfined) states and our proposed signature of deconfinement, the negative values of the strange-quark chemical potential.

The experimental observation of negative strange-quark chemical potential values in sulfur-induced interactions at 200 AGeV, together with the proposed notion that it indicates deconfinement, suggest that thermodynamic quantities of equilibrated primordial states, situated beyond the HG phase, may indeed be measured. This appears at first as 'impossible', since hadronization always takes place on the deconfinement-hadronization line, separating the HG phase from the DQM one and, therefore, we should have knowledge of these quantities only on this line. That is, always $\mu_s$ = 0 and, for the sulfur-induced interactions with $\mu_q$ ~ 82 MeV, T ~ 175 MeV [22,23].

To overcome this apparent difficulty, we propose that the conservation of the quark fugacities, $\lambda_i = \exp(\mu_i/T)$, (i = u, d, s), is a characteristic property of strong interactions and thermodynamic equilibration in general, affecting all thermally and chemically equilibrated states throughout the phase diagramme. That is, the fugacities $\lambda_i$, determining the quark number density $n_i$, once fixed in an equilibrated primordial state with finite baryon number density located beyond the HG phase, are constants of the entire sequent evolution process. This is contingent on an isentropic path and hadronization via a second order phase transition (fast hadronization without mixed phase). Both of these notions are now fairly accepted. This statement has far-reaching consequences for defining a primordial state situated in the deconfined partonic region, since the thermodynamic quantities T, $\mu_q$, $\mu_s$ of the state may indeed be determined from the experimental hadron yields and an appropriate EoS.

Thermal model fits to sulfur-induced interactions gave negative $\mu_s$ values, which is not a characteristic of the hadronic phase, as well as temperatures in the range of 180-190 MeV, which are 5-15 MeV higher than the maximum temperature for deconfinement, $T_d \sim$ 175 MeV at $\mu_q \sim$ 82 MeV, as given by the SSBM. Both of these observations point to the likelihood of our suggestions. Note also that negative strange-quark chemical potential means that the particle yield ratio: $\Omega^+/\Omega^- = \exp(-6\mu_s/T)$ is greater than one[4]. For the Pb+Pb interaction at 158 AGeV, situated in the hadronic phase ($\mu_s > 0$), it was found experimentally that $\Omega^+/\Omega^- = 0.383$, [28].

To scan the region of negative $\mu_s$ values, in a finite baryon number density DQM phase, an 'excitation function' should be performed at the Relativistic Heavy Ion Collider, RHIC at Brookhaven with Au+Au collisions at energies $20 < \sqrt{s} < 100$ AGeV. Higher energies give very small baryochemical potential at mid-rapidity[4] [29,30], hence almost zero $\mu_s$ throughout the phase diagramme. If the observation of negative $\mu_s$ values, together with temperatures in excess of $T_d$, is confirmed at these energies, it will be a profound observation, indicating that the negative strange-quark chemical potential is indeed a unique, concise and well-defined signature of the deconfinement phase transition, identifying the partonic phase. Of equal importance will be the possibility to determine the thermodynamic quantities $\mu_q$, $\mu_s$ and T, hence, the energy density and entropy of equilibrated primordial states situated far *beyond* the hadronic phase, in the deconfined-quark − gluon region.

---

[4] For Au+Au interactions at RHIC ($\sqrt{s}$ = 130 AGeV) we predict $\Omega^+/\Omega^- \sim 1$, since $\mu_s \sim 0$, for small light-quark chemical potential ($\mu_q \sim$ 14 MeV). From the experimental data compiled in ref. [32] we calculate: $\mu_q/T$ = 0.073 ± 0.006 and $\mu_s/T$ = 0.005 ± 0.01.

**Table 1.** Deduced values for T, $\mu_q$, $\mu_s$ from thermal – statistical model calculations and fits to experimental data for several nucleus-nucleus interactions.

| Interaction/Experiment | Si + Au (14.6 AGeV) E802 | | |
|---|---|---|---|
| | Ref.[33] | Ref.[26] | Mean |
| T (MeV) | 134 ± 6 | 135 ± 4 | 135 ± 3 |
| $\mu_q$ (MeV) | 176 ± 12 | 194 ± 11 | 182 ± 5 |
| $\mu_s$ (MeV) | 66 ± 10 | | 66 ± 10 |

| Interaction/Experiment | Au + Au (11.6 AGeV) E802 | | |
|---|---|---|---|
| | Ref.[33] | Ref.[26] | Mean |
| T (MeV) | 144 ± 12 | 121 ± 5 | 124 ± 5 |
| $\mu_q$ (MeV) | 193 ± 17 | 186 ± 5 | 187 ± 5 |
| $\mu_s$ (MeV) | 51 ± 14 | | 51 ± 14 |

| Interaction/Experiment | Pb + Pb (158 AGeV) NA49 | | | |
|---|---|---|---|---|
| | Ref.[33] | Ref.[26] | Ref.[27] | Mean |
| T (MeV) | 146 ± 9 | 158 ± 3 | 157 ± 4 | 157 ± 3 |
| $\mu_q$ (MeV) | 74 ± 6 | 79 ± 4 | 81 ± 7 | 78 ± 3 |
| $\mu_s$ (MeV) | 22 ± 3 | | 25 ± 4 | 23 ± 2 |

| Interaction/Experiment | Pb + Pb (40 AGeV) NA49 |
|---|---|
| | Ref.[33] |
| T (MeV) | 147 ± 3 |
| $\mu_q$ (MeV) | 136 ± 4 |
| $\mu_s$ (MeV) | 35 ± 4 |

| Interaction/Experiment | S + S (200 AGeV) NA35 | | | |
|---|---|---|---|---|
| | Ref.[33] | Ref.[25] | Ref.[24] | Mean |
| T (MeV) | <189 ± 8> | 181 ± 11 | 202 ± 13 | 189 ± 8 |
| $\mu_q$ (MeV) | 95 ± 12 | 74 ± 7 | 87 ± 7 | 83 ± 5 |
| $\mu_s$ (MeV) | - 51 ± 30 | - 58 ± 18 | | - 56 ± 15 |

| Interaction/Experiment | S + Ag (200 AGeV) NA35 | | | |
|---|---|---|---|---|
| | Ref.[33] | Ref.[25] | Ref.[24] | Mean |
| T (MeV) | <182 ± 6> | 179 ± 8 | 185 ± 8 | 182 ± 6 |
| $\mu_q$ (MeV) | 87 ± 11 | 81 ± 6 | 81 ± 7 | 82 ± 4 |
| $\mu_s$ (MeV) | - 15 ± 25 | - 65 ± 20 | | - 45 ± 16 |

**Figure Captions**

1. Variation of strange-quark chemical potential with temperature (Eq's 2,4,6) in the 3-region phase diagramme. The curve $\mu_s(T)$ intersects the $(T, \mu_q)$-plane at the intersection point of the $\mu_q = 0.45T$ line and the SSBM deconfinement line.

2. Plot of the mean values of the deduced quantities, temperature and strange-quark chemical potential, from thermal–statistical model fits to experimental data for several interactions.

3. The phase diagramme with the SSBM deconfinement line and the location of the states of nucleus-nucleus interactions in the two phases HG and DQM, deduced from the analysis of thermal models (Table 1).

4. Variation of the strange-quark chemical potential throughout the phase diagramme (Eq's 2,4,6) for $\mu_q = 82$ MeV, and the average points $(T, \mu_s)$ corresponding to the sulfur-induced interactions.

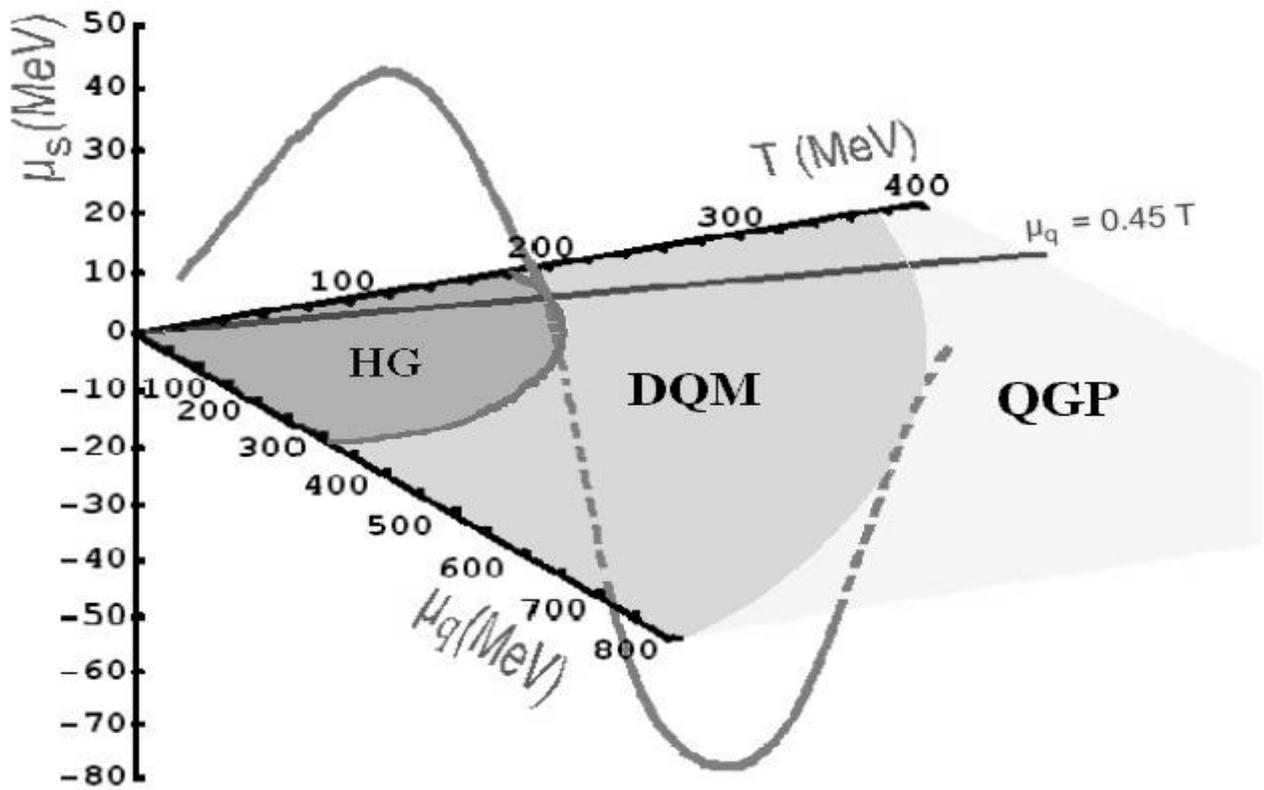

fig. 1

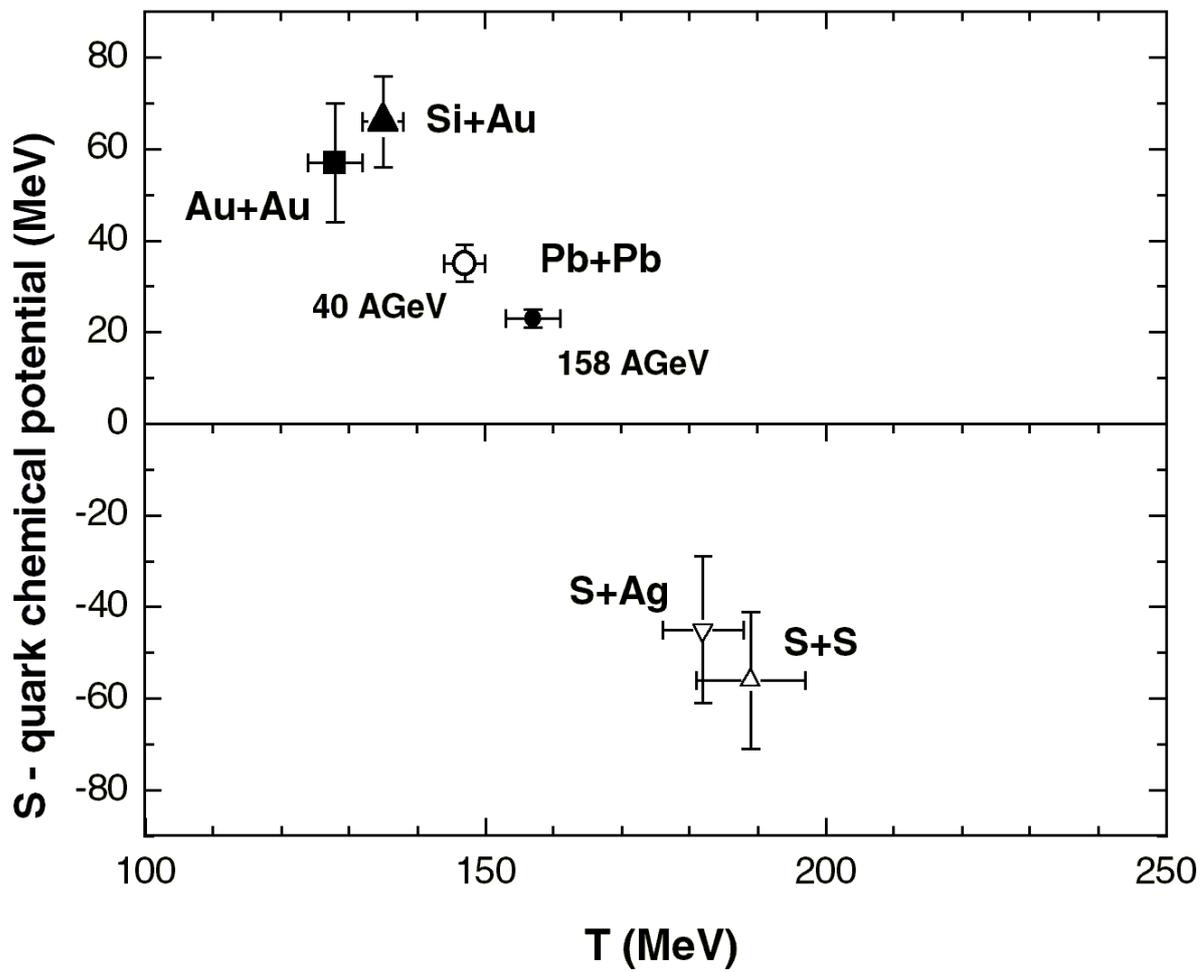

**Figure 2**

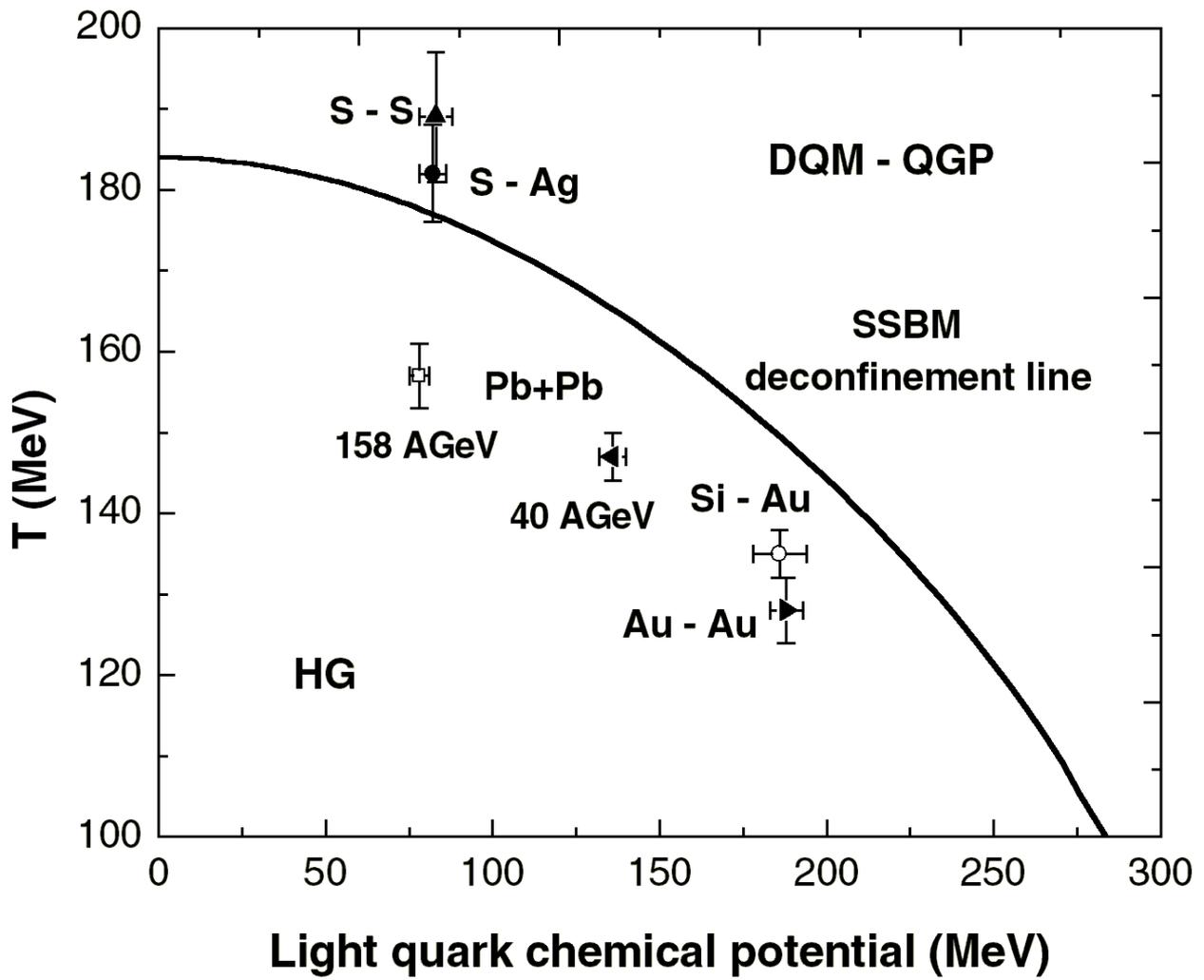

Figure 3

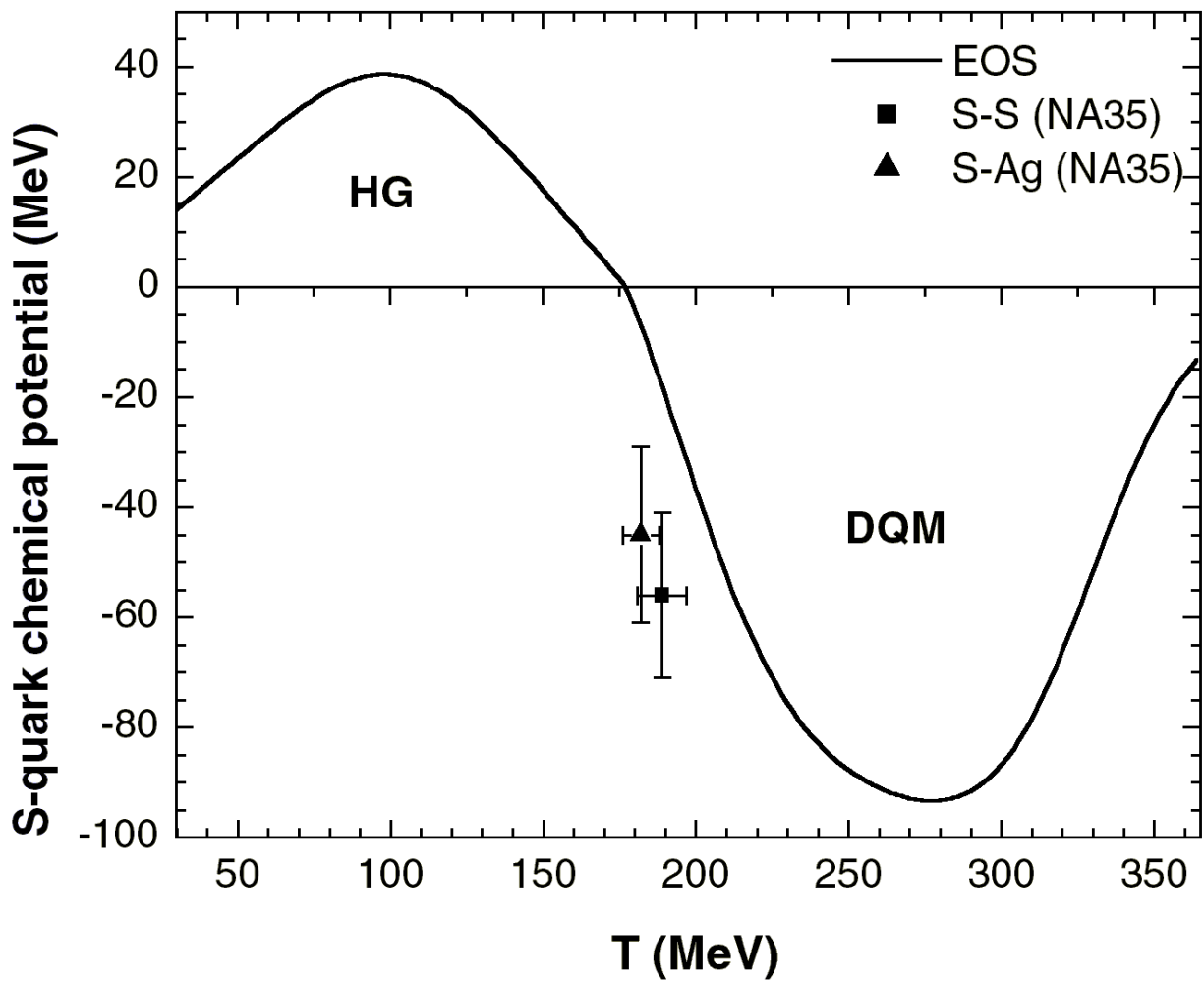

**Figure 4**